\documentstyle[twocolumn,prb,aps,epsf]{revtex} 
 
\begin{document} 
 
\twocolumn[  
\title{Correlated atomic motion, phonons and spin-ordering in bcc $^{3}$He}   
\author{N. Gov and E. Polturak}   
\address{Physics Department,\\   
Technion-Israel Institute of Technology,\\   
Haifa 32000, Israel}   
\maketitle   
\tightenlines  
\widetext  
\advance\leftskip by 57pt  
\advance\rightskip by 57pt

\begin{abstract}   
We propose a new way to treat nuclear magnetism of solid $^{3}$He.  
We argue that the magnetic interaction arises indirectly as a consequence of correlated   
zero-point motion of the ions. This motion   
lowers the energy of the ground state,   
and results in a coherent state of oscillating electric dipoles. Distortion of the electronic 
wavefunctions leads to hyperfine magnetic interactions with the nuclear spin. 
 Our model describes   
both the phonon spectra and the  
 nuclear magnetic ordering of bcc $^{3}$He using a single parameter, the  
 dipolar interaction energy $E_{0}$. The model yields correctly both the u2d2 symmetry   
of the ordered phase and the volume dependence of the magnetic interaction.  
\end{abstract}   
   
\vskip 0.2cm  
] 
 
\narrowtext  
\tightenlines  
\vspace{.2cm} 
 
The spin-ordered phase of bcc $^{3}$He presents a difficult challenge to 
accurate theoretical description \cite{fisher,cross}. The main problem is to 
explain why the transition temperature \cite{halperin} of 10$^{-3}$K is two 
orders of magnitude larger than the nuclear dipolar interaction $\sim $10$%
^{-5}$K. The prevalent description is in terms of atomic exchange cycles 
that produce competing ferro and antiferromagnetic interactions \cite{roger}%
. This model has conceptual problems that arise from the fact that in order 
to fit experimental data, many exchange cycles involving a large number of 
atoms are needed. In fact, it is not sure that this expansion converges \cite 
{cross,ceperley}. Additionally, an overall consistent description of the 
experimental data has still not been achieved \cite{fisher,grey1}. It is 
therefore of interest to consider another approach. We propose that magnetic 
ordering is due to correlations in the zero point atomic motion, which 
however does not involve exchange of atoms \cite{niremil}. These 
correlations lower the energy of the ground state \cite{nirbcc} and result 
in a coherent state of oscillating electric dipoles, which modify the phonon 
spectrum. Finally, this motion produces an oscillating magnetic polarization 
of the electronic cloud which interacts with the nuclear spin. We show below 
that this hyper-fine type interaction has the right order of magnitude, and 
leads naturally to the distinct u2d2 antiferromagnetic phase. This scenario 
is somewhat similar to the mechanism of hyperfine enhanced nuclear magnetism 
that occurs in certain insulators \cite{suzuki}. The calculated volume 
dependence of the magnetic interaction is in good agreement with 
experimental data. 
 
At temperatures which are high compared to the magnetic interactions (T$\gg $%
1mK), the local motion in bcc $^{3}$He can be treated in the same way as in $%
^{4}$He \cite{niremil}. The usual Helium inter-atomic potential is the 
Van-der Waals interaction, due to the zero-point fluctuations of the $s$%
-shell electrons. This interaction is calculated assuming that the nuclei 
are stationary (Born-Oppenheimer approximation), which is reasonable since 
the nuclear motion is usually negligible. The result of averaging over 
random electric dipolar fluctuations leads to an isotropic $1/r^{6}$ 
attraction. In addition there is the hard-core repulsion due to the overlap 
of the filled electronic shells. This interaction, strictly valid for an 
isotropic medium, is also used in anisotropic cases, such as for atoms in a 
crystal. We would like to describe the lowest correction to the isotropic 
interaction resulting from the zero point motion of the He ions. This 
motion, resulting from confinement, is {\em especially} large and 
anisotropic in solid He. In the bcc phase, this motion is large along the 
major axes (100,010,001), due to a shallow, double-minimum lattice potential  
\cite{niremil,glyde}. If we relax the Born-Oppenheimer approximation, and 
allow some relative motion between the ion and the electrons, we will obtain 
that the zero-point motion of the ion creates an oscillating electric 
dipole. One possible source of this relative motion may be the large 
difference in inertia between the ion and electrons. If this ionic motion is 
random and isotropic it would have no observable consequences. However, we 
will show that the solid phase can gain energy if this ionic motion is 
correlated. 
 
Directional oscillation of the nucleus will break the rotational symmetry of 
the nuclear position relative to the electronic cloud. This is equivalent to 
a mixing of the $s$ and $p$ electronic levels. The size of this mixing \cite 
{niremil} is of the order of the square root of the ratio of the oscillation 
frequency of the nucleus compared with the frequency of the $s\rightarrow p$ 
electronic transition, and comes out $\sim 10^{-2}$. The distortion of the 
electronic wavefunction is described through the mixing coefficient $\lambda  
$, defined by $\left| \psi \right\rangle \simeq \left| s\right\rangle 
+\lambda \left| p\right\rangle $.We therefore have in the bcc phase an 
isotropic Van-der Waals interaction with an additional dipolar interaction 
acting along the major axes. If the relative phases of the oscillatory 
motion between different ions are correlated the net interaction energy can 
be non-zero, and becomes negative for special 'antiferroelectric' 
configurations, in which the phase of the oscillations of neighboring atoms 
differs by $\pi $ (see Fig.1). This configuration has a zero total dipole 
moment, as required. The dipolar potential for the case of correlated motion 
has a long range, $1/r^{3}$ dependence \cite{cohen}. The ground-state of the 
crystal in which the zero-point motion of the atoms is correlated may be 
described as a global state of quantum resonance (coherent state \cite 
{niremil}) between the two degenerate configurations shown in Fig.1, each of 
which minimizes the dipolar interaction energy $E_{dip}$ \cite{niremil}  
\begin{equation} 
E_{dip}=\left| {\bf \mu }\right| ^{2}\sum_{i\neq 0}\left[ \frac{3\cos 
^{2}\left( {\bf \mu }\cdot \left( {\bf r}_{0}-{\bf r}_{i}\right) \right) -1}{%
\left| {\bf r}_{0}-{\bf r}_{i}\right| ^{3}}\right]  \label{edip} 
\end{equation} 
 
where ${\bf r}_{0}$ is the position of an atom and the summation ${\bf r}_{i} 
$ runs over the entire lattice. The atomic electric dipole moment $\left|  
{\bf \mu }\right| $ induced by the coherent motion is proportional to $%
\lambda $ through  
\begin{equation} 
\left| {\bf \mu }\right| =e\left\langle \psi \left| x\right| \psi 
\right\rangle \simeq 2e\lambda \left\langle s\left| x\right| p\right\rangle  
\label{mu0} 
\end{equation} 
 
We would like to describe the effect of this additional interaction on the 
phonon spectrum. Except for the T$_{1}$(110) branch, the phonons of the 
solid are well described by the Self-Consistent Harmonic (SCH) approximation  
\cite{glyde}. The additional part is given by an effective Hamiltonian \cite 
{niremil} which describes the coherent dipolar interactions. This 
Hamiltonian was used to describe the interaction between excitons in a 
dielectric medium \cite{anderson,hopfield}, where the flipping of a dipole 
out of the ground state is treated as a local mode with bare energy \cite 
{niremil} $E_{0}=2\left| E_{dip}\right| $ . The excitations of this 
Hamiltonian are a periodic modulation of the relative phases of the dipoles, 
with respect to the ground state array (Fig.1). Due to the lower symmetry of 
the dipolar array compared to the crystal symmetry a coupling between the 
usual phonons and the dipolar modulation exists only along the (110) 
direction \cite{niremil}. The hybridization of the dipolar modulation and 
the usual phonons results in a soft T$_{1}$(110) mode with approximately 
half the energy given by SCH calculation. At the edge of the Brilloin zone 
this phonon has the energy $E_{0}$. In addition there appears a localized 
excitation of energy $2E_{0}$, which is involved in mass diffusion and 
contributes to the specific-heat \cite{niremil}. The excitations in our 
picture are therefore three phonon branches (only one of which is 
renormalized) and a localized excitation which is a quantum analogue of a 
point defect. 
 
In the case of bcc $^{4}$He the bare local mode energy $E_{0}$ was measured 
directly by NMR of a $^{3}$He impurity, and the phonon spectrum was measured 
using neutron scattering. Our calculation is consistent with both sets of 
data \cite{niremil}. There are no neutron scattering measurements of the 
phonon spectrum of bcc $^{3}$He, but there are sound velocity data that 
indicate that the slope of the T$_{1}$(110) phonon is about half of the SCH 
calculation \cite{greywall,kohler}. Since we predict that this ratio should 
indeed be 0.5, we take half the energy of the calculated \cite{kohler} SCH T$%
_{1}$(110) phonon at the edge of the Brillouin zone to be the bare 
dipole-flip energy $E_{0}\simeq 5$K at V=21.5cm$^{3}$/mole. According to our 
model, the energy of the localized mode involved in thermally activated 
self-diffusion is $2E_{0}\simeq 10$K. This value is in excellent agreement 
with the activation energy measured by x-ray diffraction, ultrasonics and 
NMR experiments \cite{heald} at V=21.5cm$^{3}$/mole. A similar activation 
energy is also obtained from the excess specific heat \cite{grey}, and 
pressure measurements \cite{izumi}. The localized excitation of energy $%
2E_{0}$ is further described in Ref. [7]. 
 
We now describe the magnetic interaction arising from our model. The lowest $%
\left| p\right\rangle $ level of the He atom has the electrons in a spin $S=1 
$ state due to strong exchange interaction \cite{cohen}, of the order of 
0.25eV. In addition this level is split into 3 sublevels with $J=L+S=0,1,2$. 
The splitting, due to spin-orbit coupling \cite{cohen}, is of the order of $%
\sim 1.5$K.\ In the ground state, the $\left| p\right\rangle $ electrons 
will reside in the $^{3}P_{2}$ sublevel, with an oscillating magnetic moment  
${\bf M}_{e}$ of size $\sim \lambda m_{e}$, where $m_{e}$ is the magnetic 
moment of an electron. Because part of the magnetic moment is now in the $%
\left| p\right\rangle $ state, there appears a net uncancelled moment of 
equal size in the $\left| s\right\rangle $ component of the electronic 
wavefunction. In $^{4}$He the nuclear spin is zero, and the magnetic moment 
of the electrons has no effect. In contrast, in $^{3}$He the nuclear 
magnetic spin $I=1/2$ will interact with the oscillating electronic magnetic 
moment, mainly due to the contact term of the $\left| s\right\rangle $ 
electron at the nucleus. The magnetic interaction is of the hyper-fine type  
\cite{cohen}, and the energy associated with it, $E_{mag}$, is given by  
\begin{equation} 
E_{mag}=\left\langle -\frac{8\pi }{3}{\bf M}_{e}\cdot {\bf M}_{n}\delta 
(r)\right\rangle   \label{edipcont} 
\end{equation} 
 
where ${\bf M}_{n}$ is the nuclear magnetic moment{\em \ }and the 
calculation of the matrix element is done just as for the $\left| 
s\right\rangle $ level of the hydrogen atom \cite{cohen}. We show below that 
the maximum value of $E_{mag}/k_{B}\simeq 0.75$mK (for V=24 cm$^{3}$/mole), 
much larger than the direct nuclear dipole-dipole interaction, and is of the 
right magnitude to explain the high transition temperature of nuclear 
ordering in bcc $^{3}$He. 
 
We would like to make a quantitative calculation of the strength of the 
magnetic interaction as a function of pressure at T=0. In our model, the 
magnetic energy (\ref{edipcont}) will change with pressure due to changes of 
the electronic magnetic polarization ${\bf M}_{e}$. First, the pressure 
changes the admixture of the $\left| p\right\rangle $ orbital in the ground 
state wavefunction $\left| \psi \right\rangle $, through changes of the 
mixing parameter $\lambda $. Second, as the solid is compressed, the 3 
sub-levels ($J=0,1,2$) broaden into partially overlapping bands. This 
overlap results in a reduction of the net alignment of the electronic 
magnetic moment ${\bf M}_{e}$. 
 
The mixing parameter $\lambda $ depends on pressure through $\sqrt{VE_{0}}$, 
where $V$ is the molar volume. This behaviour arises from the definition of $%
E_{dip}$ (\ref{edip}),(\ref{mu0}) and the factor $\left| {\bf r}_{0}-{\bf r}%
_{i}\right| ^{3}$ in this equation is proportional to the molar volume $V$. 
We take $E_{dip}$ as a function of pressure from the measured \cite 
{heald,greywall} activation energy $2E_{0}$. $\lambda $ is found to increase 
with decreasing molar volume $V$. The effect of the broadening of the 
sub-levels with pressure can be approximated using the overlap integral of 
the three sub-levels  
\begin{equation} 
\left\langle {\bf M}_{e}(V)\right\rangle \simeq \left( 1-F(V)\right) \lambda 
(V){\bf m}_{e}  \label{magf} 
\end{equation} 
 
where the overlap integral $F(V)=\int \psi _{1}^{\ast }(E)\psi _{-1}(E)dE$. 
Here, $\psi _{1}^{\ast }(E)$ and $\psi _{-1}(E)$ are taken as normalized 
gaussians centered at the energies of the spin-orbit aligned and 
anti-aligned levels, which are $\sim 1.5$K apart \cite{cohen}. The effective 
energy width $\gamma $ of $\psi _{1}^{\ast }(E)$ and $\psi _{-1}(E)$ is 
estimated using simple band calculation, as the change of the Coulomb energy 
of the $\left| p\right\rangle $ electrons due to overlaping $\left| 
p\right\rangle $ wavefunctions on neighboring He nuclei \cite{ashcroft}  
\begin{equation} 
\gamma (V)=\lambda ^{2}\sum_{i}\int \frac{e^{2}}{\left| {\bf r}_{0}-{\bf r}%
_{i}\right| }\psi _{0}^{\ast }(r)\psi _{i}(r)d^{3}r  \label{width} 
\end{equation} 
 
where $\psi _{0}^{\ast }(r),\psi _{i}(r)$ are the $p$-state wavefunctions of 
the electrons around the central atom and its $i$ neighbors. In this 
calculation we also take into account the large spatial spread in the atomic 
position inside the wide potential well \cite{niremil}, of approximately $%
\pm 0.8$\AA . We find that $\gamma $ ranges from $\sim $0K for V=24 cm$^{3}$%
/mole to 5.5K for V=19 cm$^{3}$/mole. The overlap factor $1-F(V)$ is very 
sensitive to volume, changing from 1 at V=24 cm$^{3}$/mole to $\sim 0.01$ at 
V=19 cm$^{3}$/mole. One can see that as the volume decreases the broadening 
of the bands increases thereby decreasing the net magnetic polarization of 
the oscillating electronic cloud. 
 
The strength of the magnetic interaction should be proportional to the 
measured Curie-Weiss temperature $\theta $. In Fig.2 we compare the magnetic 
splitting $2E_{mag}$ (\ref{edipcont}) with the two available sets of 
measured \cite{fisher} $\theta $. We find that the volume dependence of $%
E_{mag}$ agrees very well with that of $\theta $, i.e. $2E_{mag}=const \cdot 
\theta$. The constant is 1 for the data in Fig.2a, and 2 for the data in 
Fig.2b. 
 
We now consider the symmetry of the ordered spin system. The existence of 
the hyper-fine splitting means that the simple quantum resonance condition 
on each site is broken. Now, the two antiferoelectric configurations shown 
in Fig. 1 are not degenerate, with an energy difference of $2E_{mag}$ per 
site. It is possible to restore the degeneracy of the overall ground state, 
and hence the quantum resonance condition. This can be done if the nuclear 
spins become ordered as well, in a spatial configuration which has the same 
interaction energy with each of the two states of Fig. 1. The electric 
dipolar interactions in the ground-state of Fig.1 are such that each simple 
cubic sublattice of the bcc has no net interaction with the other 
sublattice. Two possible static arrangements of the nuclear spins that 
fulfill the resonance requirement on each sublattice are shown in Fig.3. 
Both arrangements are the simplest which ensure that there is an equal 
number of atoms with electronic and nuclear spins aligned (and anti-aligned) 
in both degenerate configurations of the electric dipoles. These 
arrangements preserve the overall time-reversal symmetry of the system at 
zero field. Since these configurations exist on both sublattices, we end up 
with an u2d2 arrangements which is the symmetry of the ordered nuclear phase  
\cite{osheroff,bossy}. Spontaneous symmetry breaking due to ordering of the 
spins will select one of the major axes (100,010,001), similar to a ferro or 
antiferromagnet. The electric dipoles oscillating along the two orthogonal 
axes will have no magnetic interaction energy, and are therefore in the 
usual quantum resonance of Fig.1. We point out that in our model, the u2d2 
symmetry of the spin ordered nuclear phase is a natural consequence of the 
coherent-state of the oscillating dipoles (Fig.1), and does not depend on 
the magnitude of the magnetic interaction. 
 
To conclude, our model enables us to describe both the phonon spectra and 
the nuclear magnetic ordering of bcc $^{3}$He using a single parameter, 
which is the thermal activation energy $E_{0}$. Experimental values of $E_{0} 
$ were measured in several experiments. The model describes correctly both 
the symmetry of the ordered phase and the volume dependence of the magnetic 
interaction. 
 
{\bf Acknowledgements} 
 
This work was supported by the Israel Science Foundation and by the Technion 
VPR fund for the Promotion of Research.

\begin{figure}[tbp] 
\input epsf \centerline{\ \epsfysize 5cm \epsfbox{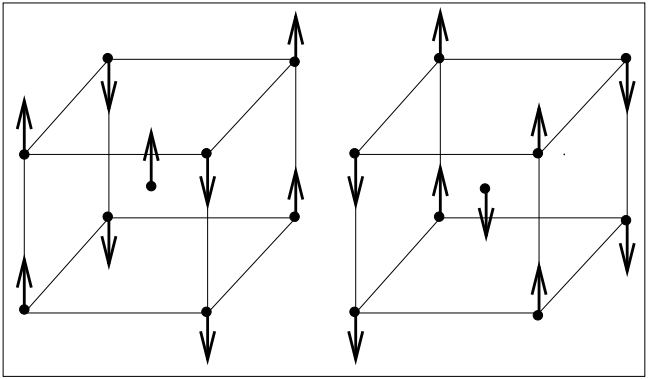}}
\vskip 3mm
\caption{The two degenerate 'antiferroelectric' dipole arrangement in the 
ground-state of the bcc phase. The arrows show the instantaneous direction 
of the dipoles.} 
\end{figure} 
 
\begin{figure}[tbp] 
\input epsf \centerline{\ \epsfysize 10cm \epsfbox{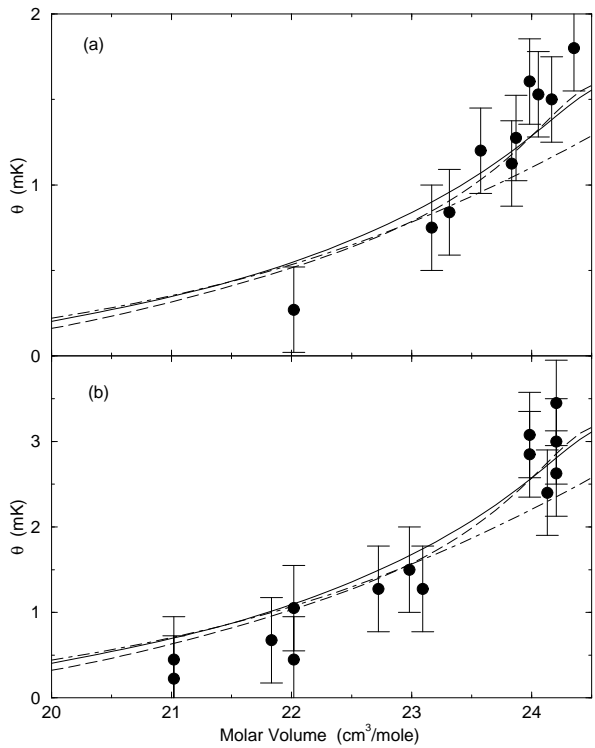}}
\vskip 3mm
\caption{The magnetic splitting $2E_{mag}$ (Eq.3) as a function of volume, 
calculated using different experimental data sets for the activation energy $%
E_0$: dashed line [18], dash-dot [17] and solid line [16]. The solid circles 
in (a) and (b) are the experimental Curie-Weiss temperature $\protect\theta$ 
[1]. To compare with the higher set of data (b), we multiplied our 
calculated splitting by an extra factor of 2.} 
\end{figure} 
 
\begin{figure}[tbp] 
\input epsf \centerline{\ \epsfysize 5cm \epsfbox{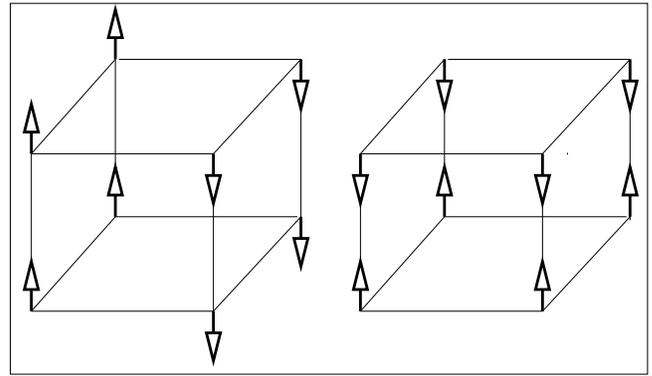}}
\vskip 3mm
\caption{The two simplest u2d2 static nuclear spin (arrows) arrangements 
that maintain the quantum resonance of both the electric and magnetic 
dipoles (Fig.1). Only the spins on one simple-cubic sublattice of the bcc 
are shown, with the other sublattice having identical arrangement.} 
\end{figure} 
 
\end{document}